\documentclass[lettersize,journal]{IEEEtran}

\usepackage{amsmath,amsfonts}
\usepackage{algorithmic}
\usepackage{array}
\usepackage[caption=false,font=normalsize,labelfont=sf,textfont=sf]{subfig}
\usepackage{textcomp}
\usepackage{stfloats}
\usepackage{url}
\usepackage{verbatim}
\usepackage{graphicx}
\def\BibTeX{{\rm B\kern-.05em{\sc i\kern-.025em b}\kern-.08em
    T\kern-.1667em\lower.7ex\hbox{E}\kern-.125emX}}
\usepackage{balance}

\newcommand{\ddt}[1]{\frac{\mathrm{d}{#1}}{\mathrm{dt}}} 
\newcommand{\infnorm}[1]{\vert\vert{#1}\vert\vert_{\infty}} 

\begin{document}

\title{DAE Index Reduction for Electromagnetic Transient Models}

\author{
\IEEEauthorblockN{
    Fiona Majeau\IEEEauthorrefmark{1}\IEEEauthorrefmark{2},
    Jose Daniel Lara\IEEEauthorrefmark{3},
    Eduardo Corona\IEEEauthorrefmark{4}, and
    Bri-Mathias Hodge\IEEEauthorrefmark{1}\IEEEauthorrefmark{2}\IEEEauthorrefmark{3}\IEEEauthorrefmark{4}
    } \\
\IEEEauthorblockA{\IEEEauthorrefmark{1} 
    Electrical, Computer and Energy Engineering,
    University of Colorado Boulder,
    United States
    } \\
\IEEEauthorblockA{\IEEEauthorrefmark{2}
    Renewable and Sustainable Energy Institute, 
    University of Colorado Boulder,
    United States
    } \\
\IEEEauthorblockA{\IEEEauthorrefmark{3} 
    Grid Planning and Analysis Center,
    National Laboratory of the Rockies, 
    Golden, United States
    } \\
\IEEEauthorblockA{\IEEEauthorrefmark{4} 
    Applied Mathematics,
    University of Colorado Boulder,
    United States
    } 
}


\markboth{PREPRINT}{} 

\maketitle

\begin{abstract}
Electromagnetic transient (EMT) models are index-2 differential-algebraic equations when they include certain topologies and are formulated with modified nodal analysis. Such systems are difficult to numerically integrate, a challenge that is currently addressed by applying model approximations or reformulating with index-reduction algorithms. These algorithms exist in general-purpose software tools, but their reliance on symbolic representation makes them computationally prohibitive for large network-wide EMT models. This paper derives and presents two modular index-reduced subsystem models that allow EMT models to be integrated with standard solvers, without approximations or symbolic algorithms. Both subsystems include a transformer, one isolated and one machine‑coupled. We measure the computational performance of constructing EMT models with up to 1152 buses using the custom subsystem models and the symbolic algorithms. The custom approach reduces memory usage and runtime of model construction by several orders of magnitude compared to the general approach, shifting the bottleneck from construction to integration.
\end{abstract}


\section{Introduction}

\IEEEPARstart{E}{lectromagnetic} transient (EMT) modeling and simulation is becoming increasingly necessary for predicting network-wide stability as more power electronic devices connect to the grid. Quasi-static phasor (QSP) modeling has been the standard approach for network-wide stability predictions, but it neglects the electromagnetic dynamics of machine stators, transformers, and lines, which we will refer to collectively as network dynamics. As electromagnetic devices become more prevalent and less localized, it is harder to claim that time-scale separation justifies the exclusion of network dynamics \cite{milanoFoundationsChallengesLowInertia2018,markovicUnderstandingSmallSignalStability2021, henriquez-aubaGridFormingInverter2020}. Since EMT models include these dynamics, there is increasing interest in running network-wide EMT simulations. Unfortunately, EMT simulation tools were originally designed to simulate small subsets of a grid, and they do not scale well for simulations of realistically-sized grids, which can have over 100,000 state variables \cite{subediReviewMethodsAccelerate2021}. Therefore, researchers are trying to improve the computational feasibility of network-wide EMT simulations.

One promising approach is a model formulation method called EMT-\textit{dq}, proposed in \cite{laraRevisitingPowerSystems2024}. This method defines the balanced network equations with respect to a rotating \textit{dq} reference frame (RF) instead of a stationary three-phase \textit{abc} RF. At a stable operating point, network states expressed in a \textit{dq} RF are constant (time-invariant) while those expressed in an \textit{abc} RF are sinusoidal (time-variant). When a system of differential equations is time-invariant, adaptive time stepping can meaningfully improve time-integration efficiency \cite{laraRevisitingPowerSystems2024}. 

To advance the development of EMT-\textit{dq} simulation tools, several open questions must be addressed. Most EMT tools use a systematic model formulation method, usually modified nodal analysis (MNA), to convert the network topology and device models into a system of differential-algebraic equations (DAEs) \cite{chung-wenhoModifiedNodalApproach1975}. However, when MNA is applied to models with network dynamics, specific network structures cause the resulting system of equations to be a numerically sensitive form called an \textit{index-2 DAE}, which is challenging to integrate \cite{tischendorfTopologicalIndexcalculationDAEs1998,ascherComputerMethodsOrdinary1998}. This paper focuses on two common structures that cause this issue: an equivalent circuit of a transformer with nonzero magnetizing reactance, and the interface between a synchronous generator (SG) stator and transformer. We will refer to these structures as S1 and S2, respectively. Typically, the transformer in S2 is an S1 transformer, so we present models for S1 and the combined subsystem, S1+S2.

For index-2 DAEs, it is generally recommended to perform a reformulation process called index reduction to achieve a form that can be reliably integrated \cite{ascherComputerMethodsOrdinary1998,brenanNumericalSolutionInitialValue1995}. DAE index reduction algorithms have been implemented in many software tools, like Dymola (Modelica), Simscape (MATLAB), and ModelingToolkit (Julia). These algorithms perform symbolic transformations on a system of equations, like differentiation and equation sorting, to produce a symbolic model. To run a numerical simulation, the symbolic model is converted into numerical code through a process called \textit{code generation}. There has been an effort to develop Modelica-based EMT simulation tools which use Modelica's symbolic transformation toolkit to perform index reduction. However, code generation is identified in \cite{guironnetOpenSourceSolutionUsing2018} as a significant barrier to building large-scale Modelica-based EMT models. We encountered the same barrier when using ModelingToolkit, as shown in  Section~\ref{sec:experimental_results}.

There is limited mention of the existence or difficulty of index-2 EMT models in the literature. S2 is briefly considered in the appendix of \cite{allenInteractionTransmissionNetwork2000}. The authors show the steps for generating a state-space representation of this structure, which successfully reduces the index. However, this is not the primary focus of their paper, so they do not extend the process to other topologies or perform numerical experiments illustrating the validity or computational performance of the method. Index-2 structures are also discussed in \cite{kulaszaExtendingFrequencyBandwidth2022}, where they are called floating subnetworks. The authors experience convergence issues when integrating a system with S2 using the DAE solver, IDA \cite{hindmarshSUNDIALSSuiteNonlinear2005}. They address this by adding a small shunt capacitor to every floating bus, which prevents convergence failures in their particular experiments and technically reduces the index. However, convergence failures could still occur for other cases because the Newton iteration matrix (see Section~\ref{sec:numerical_integration}) is only slightly less ill-conditioned than the original system due to the small capacitor values.

S1 and S2 also cause EMT-\textit{abc} models to be index-2 DAEs. The PSCAD user guide provides insight on how both structures are handled \cite{EMTDCUsersGuide2010}. S1 is formulated from Maxwell's equations instead of the equivalent circuit, which generates an index-1 model. S2 is approximated by calculating the stator current injection in between two time-integration steps of the decoupled network solve. However, the stator current and terminal bus voltage are dynamically coupled and should be updated within the same time step. The user guide reports spurious voltage spikes at these simulated interfaces when the voltage changes quickly, which is likely an artifact of this decoupling. To compensate, the developers added a shunt resistor and current source to the interface model.

The main contribution of this paper is a set of modular index-reduced subsystem models that resolve a numerical challenge caused by common components, enabling large network-wide EMT-\textit{dq} simulations. EMT-\textit{dq} models with these subsystems are compatible with standard index-1 DAE or ODE solvers. Without these subsystem models, there are only two options to reliably generate time-domain simulations of index-2 EMT-\textit{dq} models: simplifying the model with approximations such as in \cite{kulaszaExtendingFrequencyBandwidth2022} and \cite{EMTDCUsersGuide2010}, or building the model with general-purpose index reduction tools. The models presented in this paper require no additional approximations and can be built using orders of magnitude less memory and time than general-purpose tools. This paper does not provide models for every possible case of index-2 DAEs in power systems, nor does it claim to present a universal algorithm to derive them. Rather, we focus on the specific application of network-wide EMT-\textit{dq} models by addressing two commonly encountered subsystems that cause them to be index-2 DAEs. 

This paper is organized as follows. Section~\ref{sec:emt_models_as_index2_daes} defines DAEs and why they show up in EMT models. Section~\ref{sec:index_reduction} discusses the numerical challenges of index‑2 DAEs, the process of index reduction, and how index reduction can be applied to EMT models. Section~\ref{sec:subsystem_model_derivations} derives two custom subsystem models and Section~\ref{sec:experimental_design} describes the numerical experiments. Section~\ref{sec:experimental_results} presents experimental results which illustrate the validity, efficiency, and scalability of the custom approach. Section~\ref{sec:conclusion} presents conclusions and future work. 

\textit{Notation}: Bold letters $\mathbf{x}$ represent column vectors of variables or functions; $\dot{\mathbf{x}}$ or $\ddt{\mathbf{x}}$ represent column vectors of element-wise time derivatives; subscripts $RI$ and $dq$ represent quantities defined with respect to the network RF and generator RF, respectively; the network RF rotates at a constant system base frequency, $\omega_0$; the generator RF for an SG rotates at the shaft frequency, $\omega$; pairs of $RI$ or $dq$ variables or functions may be represented as column vectors, e.g. $\mathbf{i}_{1,RI} = [i_{1,R}, i_{1,I}]^T$ or $\mathbf{i}_{1,RI} \pm \mathbf{i}_{1,IR}= [i_{1,R} + i_{1,I}, \, i_{1,I} - i_{1,R}]^T$; quantities are in per unit (pu) unless otherwise specified; the system is assumed to be balanced, so 0-axis equations are excluded.

\section{EMT Models as Index-2 DAEs}
\label{sec:emt_models_as_index2_daes}

\subsection{Definitions}
\label{sec:definitions}
Consider a nonlinear system of DAEs in semi-explicit form, given by \eqref{eqn:dae}. The following discussion is restricted to this class of problems, since EMT-\textit{dq} models are typically expressed in this form. We refer to elements of $\mathbf{x}$ as \textit{differential variables} and elements of $\mathbf{z}$ as \textit{algebraic variables}. The Jacobian and implicit form of \eqref{eqn:dae} are given by \eqref{eqn:dae_jacobian} and \eqref{eqn:dae_implicit}, respectively. We assume that \eqref{eqn:dae} does not include redundant equations. 

\vspace{-3mm}
\begin{minipage}[t]{0.38\linewidth}
\vspace{0pt}
\begin{subequations}
\begin{align}
    \dot{\mathbf{x}} &= \mathbf{f}(t, \mathbf{x}, \mathbf{z})  \label{eqn:dae_f}
    \\
    \mathbf{0} &= \mathbf{g}(t, \mathbf{x}, \mathbf{z}) \label{eqn:dae_g}
\end{align}\label{eqn:dae}
\end{subequations}
\end{minipage}
\begin{minipage}[t]{0.02\linewidth}
\ 
\end{minipage}
\begin{minipage}[t]{0.55\linewidth}
\vspace{6pt}
\begin{align}
    \mathbf{J}\left(t, \mathbf{x}, \mathbf{z} \right) = 
    \begin{bmatrix}
        \mathbf{f}_{\mathbf{x}} & \mathbf{f}_{\mathbf{z}}  \\ \mathbf{g}_{\mathbf{x}} & \mathbf{g}_{\mathbf{z}}
    \end{bmatrix} 
    \label{eqn:dae_jacobian}
\end{align}
\end{minipage}
\vspace{-10pt}
\begin{subequations}
    \begin{align}
    \mathbf{F}\left(t,\mathbf{x}, \mathbf{z}, \dot{\mathbf{x}} \right) &= \mathbf{f}(t, \mathbf{x}, \mathbf{z}) - \dot{\mathbf{x}}
    \\
    \mathbf{G}\left(t,\mathbf{x}, \mathbf{z}, \dot{\mathbf{x}} \right) &= \mathbf{g}(t, \mathbf{x}, \mathbf{z})
\end{align} \label{eqn:dae_implicit}
\end{subequations}

The \textit{incidence matrix} of \eqref{eqn:dae} is given by \eqref{eqn:dae_incidence}, where the operator $\mathcal{S}\{\mathbf{A}\}$ denotes the sparsity pattern of matrix $\mathbf{A}$. The incidence matrix, $\mathbf{B}$, is a binary sparsity pattern matrix that defines the structural dependence of the system on $\dot{\mathbf{x}}$ and $\mathbf{z}$. It is related to $\mathbf{J}$ through the sparsity patterns of $\mathbf{f}_{\mathbf{z}}$ and $\mathbf{g}_{\mathbf{z}}$ \cite{cellierContinuousSystemSimulation2006}. 
\begin{align}
    \mathbf{B} &= 
    \begin{bmatrix}
        \mathcal{S}\{ \frac{\partial{\mathbf{F}}}{\partial \dot{\mathbf{x}}} \} & 
        \mathcal{S}\{\frac{\partial{\mathbf{F}}}{\partial{\mathbf{z}}}\} \\[1ex] 
        \mathcal{S}\{\frac{\partial{\mathbf{G}}}{\partial \dot{\mathbf{x}}}\} & 
        \mathcal{S}\{\frac{\partial{\mathbf{G}}}{\partial{\mathbf{z}}}\} \\ 
    \end{bmatrix}
    =
    \begin{bmatrix}
        \mathbf{I} & \mathcal{S}\{\mathbf{f}_{\mathbf{z}}\} \\ 
        \mathbf{0} & \mathcal{S}\{\mathbf{g}_{\mathbf{z}}\}
    \end{bmatrix} 
    \label{eqn:dae_incidence}
\end{align}
\begin{align*}
    \text{where \quad} 
    \mathcal{S}\{\mathbf{A}\} =
    \begin{cases}
        1 & \mathbf{A}_{ij} \neq 0\\
        0 & \mathbf{A}_{ij} = 0
    \end{cases} \nonumber 
\end{align*}

A system of DAEs can be categorized with an integer called the \textit{index}, which describes how it relates to an equivalent system of ODEs. The index of \eqref{eqn:dae} is the minimum number of times that \eqref{eqn:dae_g} must be differentiated with respect to $t$ to determine $\dot{\mathbf{z}}$ as a continuous function of $t,\mathbf{x},$ and $\mathbf{z}$ \cite{brenanNumericalSolutionInitialValue1995}. An index-0 DAE is an ODE. A semi-explicit system of DAEs given by \eqref{eqn:dae} and \eqref{eqn:dae_jacobian} is index-1 if and only if $\mathbf{g}_{\mathbf{z}}$ is non-singular. Otherwise, it is index-2 or higher, a category referred to as \textit{higher-index DAEs} \cite{brenanNumericalSolutionInitialValue1995}. 

Converting a higher index DAE to a lower index DAE is called \textit{index reduction}. Differentiating \eqref{eqn:dae_g} once and solving for $\dot{\mathbf{z}}$ gives \eqref{eqn:ddtz}. If $\mathbf{g}_{\mathbf{z}}$ is non-singular, this expression can be evaluated, indicating that \eqref{eqn:dae} is index-1 and \eqref{eqn:ddtz} is index-0. If $\mathbf{g}_{\mathbf{z}}$ is singular, then \eqref{eqn:dae} is index-2 or higher, and subsequent differentiations are required to reduce the index \cite{brenanNumericalSolutionInitialValue1995}.
\begin{subequations}
    \begin{align}
        \dot{\mathbf{x}} &= \mathbf{f}(t, \mathbf{x}, \mathbf{z}) 
        \label{{eqn:ddtz_f}}\\
        \dot{\mathbf{z}} &= \overline{\mathbf{g}}(t,\mathbf{x}, \mathbf{z}) 
        = - \mathbf{g}_\mathbf{z}^{-1} \left(\frac{\partial{\mathbf{g}}}{\partial{t}} + \mathbf{g}_\mathbf{x} \, \mathbf{f}(t, \mathbf{x}, \mathbf{z}) \right)
        \label{eqn:ddtz_g}
    \end{align}
    \label{eqn:ddtz}
\end{subequations}
Only a specific subset of $\mathbf{g}$, called \textit{constraint equations}, must be differentiated to reduce higher-index DAEs. They represent mutual dependencies between elements of $\mathbf{x}$ and can be identified from the structure of $\mathbf{g}_{\mathbf{z}}$. Consider the case where row $i$ of $\mathbf{g}_{\mathbf{z}}$ is all zeros. Since $\mathbf{g}_i$ is only a function of $\mathbf{x}$, it is a constraint equation. The corresponding elements of $\mathbf{\dot{x}}$ are also mutually dependent through $\ddt{}\mathbf{g}_i$, which is called a \textit{hidden constraint equation}. Consider the case where rows $j$ and $k$ of $\mathbf{g}_{\mathbf{z}}$ are nonzero and linearly dependent. The corresponding constraint equation can be formed by combining $\mathbf{g}_j$ and $\mathbf{g}_k$ through $\mathbf{z}$. The number of constraint equations in \eqref{eqn:dae} is the nullity of $\mathbf{g}_{\mathbf{z}}$. However, most algorithms for identifying constraint equations are performed on $\mathbf{B}$, which only represents the sparsity pattern of $\mathbf{g}_{\mathbf{z}}$. Therefore, only the zero-row constraint equations can be preemptively detected and differentiated. For the systems in this paper, all constraint equations appear as rows of zeros.

If a DAE is sufficiently differentiable, the index reduction process does not simplify or approximate the relationships represented by the equations. For example, if $\mathbf{g}_{\mathbf{z}}$ is non-singular, then \eqref{eqn:dae} and \eqref{eqn:ddtz} model the same dynamical system. However, if both models are numerically integrated with the same initial condition to get $\mathbf{x}(t)$ and $\mathbf{z}(t)$, the solution of \eqref{eqn:ddtz} may drift away from the solution of \eqref{eqn:dae}. Since \eqref{eqn:ddtz} does not explicitly include \eqref{eqn:dae_g}, the solver will not enforce those relationships and rounding errors will propagate \cite{brenanNumericalSolutionInitialValue1995}. In practice, \eqref{eqn:ddtz_g} is added to the system and \eqref{eqn:dae_g} is retained. 

\subsection{Model formulation meets topology}
\label{sec:model_formulation}
Model formulation methods are procedures for identifying the relevant variables and relationships in a physical system and expressing them mathematically. The DAE index of a model is influenced by both the modeled system and the formulation method. There are several contexts in which DAE models arise from formulation methods. When a model is formulated with conservation laws, such as conservation of mass or momentum, algebraic relationships between variables are introduced. The EMT models discussed in this paper are DAEs due to Kirchhoff's Current Law (KCL) and Kirchhoff's Voltage Law (KVL) which enforce conservation of charge and energy, respectively. Algebraic equations also emerge when stiff systems of ordinary differential equations (ODEs) are approximated using singular perturbation theory. 

\begin{figure}[!ht]
    \centering
    \begin{minipage}{.235\textwidth}
        \centering
        \includegraphics[width=0.95\textwidth]{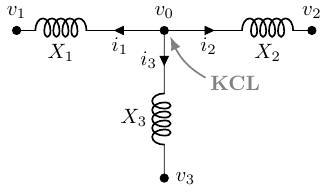}
        \caption{Example of a cutset of inductors or current sources. KCL at $v_0$ introduces constraint equations.}
       \label{fig:current_cutset}
    \end{minipage}\hfill%
    \begin{minipage}{0.235\textwidth}
        \centering
        \includegraphics[width=0.99\textwidth]{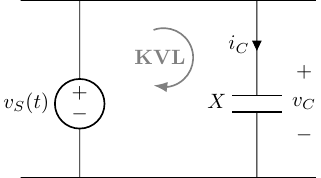}
        \caption{Example of a loop of capacitors and voltage sources. KVL introduces constraint equations.}
        \label{fig:voltage_loop}
    \end{minipage}
\end{figure}

A common formulation method for dynamic power system models is MNA \cite{chung-wenhoModifiedNodalApproach1975}. When network dynamics are retained, there are two structures of passive components that introduce constraint equations: a cutset of inductors and/or current sources (e.g. Fig.~\ref{fig:current_cutset}), and a loop of capacitors and voltage sources (e.g. Fig.~\ref{fig:voltage_loop}) \cite{tischendorfTopologicalIndexcalculationDAEs1998}. When MNA is applied to a system with one of these structures, the system of equations is an index-2 DAE. This result is an artifact of the physical system topology, the decision to model inductor and capacitor dynamics, and the model formulation method. This result applies to both EMT-\textit{abc} and EMT-\textit{dq} models, but we focus on the EMT-\textit{dq} case. The next two sections show why applying MNA to these topologies generates constraint equations.

\vspace{-4mm}
\subsection{Constraint equations from KCL}
Consider a network that includes an instance of Fig.~\ref{fig:current_cutset}. When formulating an EMT-\textit{dq} model of this network using MNA, this structure contributes three pairs of d-axis and q-axis inductor equations, \eqref{eqn:cc_i1dq}-\eqref{eqn:cc_i3dq}, and KCL at the central node, \eqref{eqn:cc_kcld}-\eqref{eqn:cc_kclq} \cite{milanoEigenvalueProblemsPower2020}. This is a DAE of form \eqref{eqn:dae} where \eqref{eqn:cc_i1dq}-\eqref{eqn:cc_i3dq} correspond to \eqref{eqn:dae_f} and \eqref{eqn:cc_kcld}-\eqref{eqn:cc_kclq} correspond to \eqref{eqn:dae_g}. 
\begin{subequations} 
    \label{eqn:cc}
    \begin{align}
        \frac{X_{1}}{\omega_0} \, \ddt{\mathbf{i}_{1,dq}} &= \mathbf{v}_{0,dq} - \mathbf{v}_{1,dq} \pm X_{1} \mathbf{i}_{1,qd} \label{eqn:cc_i1dq}\\
        \frac{X_{2}}{\omega_0} \, \ddt{\mathbf{i}_{2,dq}} &= \mathbf{v}_{0,dq} -\mathbf{v}_{2,dq} \pm X_{2} \mathbf{i}_{2,qd} \label{eqn:cc_i2dq}\\
        \frac{X_{3}}{\omega_0} \, \ddt{\mathbf{i}_{3,dq}} &= \mathbf{v}_{0,dq} -\mathbf{v}_{3,dq} \pm X_{3} \mathbf{i}_{3,qd} \label{eqn:cc_i3dq}\\
        0 &= i_{1,d} + i_{2,d} + i_{3,d} 
        \label{eqn:cc_kcld} \\
        0 &= i_{1,q} + i_{2,q} + i_{3,q}
        \label{eqn:cc_kclq} 
    \end{align}
\end{subequations}
\vspace{-5mm}
\begin{align*}
    \qquad  \text{where} \quad \mathbf{x} = \{ \mathbf{i}_{1,dq},\mathbf{i}_{2,dq},\mathbf{i}_{3,dq} \}, \,\, 
    \mathbf{z} = \{ \mathbf{v}_{0,dq} \} 
\end{align*}

Since \eqref{eqn:cc_kcld}-\eqref{eqn:cc_kclq} only include elements of $\mathbf{x}$, they are constraint equations. Only two of the three currents in \eqref{eqn:cc_kcld} must be known to calculate the third and their time derivatives are similarly constrained. The time evolution of one current is therefore defined by both its differential equation and the hidden constraint equation. The same logic applies to \eqref{eqn:cc_kclq}. Any EMT model with this structure, such as S1 and S2, is an index-2 DAE when formulated with MNA.

\subsection{Constraint equations from KVL}
Next, consider a network that includes an instance of Fig.~\ref{fig:voltage_loop}. The voltage source, defined by $F_d(t)$ and $F_q(t)$, is an independent input, not an element of $\mathbf{x}$ or $\mathbf{z}$. When formulating an EMT-\textit{dq} model of this network using MNA, this structure will contribute one pair of d-axis and q-axis capacitor equations, \eqref{eqn:vl_vdq}, and expressions of the independent voltage source, \eqref{eqn:vl_sd}-\eqref{eqn:vl_sq}. This is a DAE of form \eqref{eqn:dae} where \eqref{eqn:vl_vdq} correspond to \eqref{eqn:dae_f} and \eqref{eqn:vl_sd}-\eqref{eqn:vl_sq} correspond to \eqref{eqn:dae_g}. 
\begin{subequations} 
\label{eqn:vl}
\begin{align}
    \frac{1}{X \omega_0} \, \ddt{\mathbf{v}_{C,dq}} &= \mathbf{i}_{C,dq} \pm \frac{1}{X} \mathbf{v}_{C,qd} 
    \label{eqn:vl_vdq}\\
    0 &= F_d(t) - v_{C,d}(t) 
    \label{eqn:vl_sd} \\
    0 &= F_q(t) - v_{C,q}(t)
    \label{eqn:vl_sq} 
\end{align}
\end{subequations}
\begin{align*}
    \qquad  \text{where} \quad 
    \mathbf{x} = \{ \mathbf{v}_{C,dq} \}, \,\, 
    \mathbf{z} = \{ \mathbf{i}_{C,dq} \} 
\end{align*}

Since \eqref{eqn:vl_sd}-\eqref{eqn:vl_sq} only include elements of $\mathbf{x}$, they are constraint equations. The time evolution of $v_{C,d}$ is defined by both its differential equation and the derivative of \eqref{eqn:vl_sd}. The same logic applies to $v_{C,q}$. Any EMT model with this structure is an index-2 DAE when formulated with MNA. 

\section{Index Reduction}
\label{sec:index_reduction}

\subsection{Difficulties of integrating of higher-index DAEs}
\label{sec:numerical_integration}
EMT models must be numerically integrated over time to predict dynamic stability. There are many methods designed for ODEs and semi-explicit index-1 DAEs. IDA, a solver based on backward differentiation formula (BDF) methods, is a common choice for implicit index-1 DAEs \cite{hindmarshSUNDIALSSuiteNonlinear2005}. 
While there are stable and consistent methods for some higher-index DAEs, notably Hessenberg index-2 DAEs\footnote{The systems in Section~\ref{sec:subsystem_model_derivations} can be cast into index-2 Hessenberg form.}, most have no standard options \cite{ascherComputerMethodsOrdinary1998}. Higher-index DAEs of any form are difficult to integrate because $\mathbf{g}_{\mathbf{z}}$ is singular, causing the Newton iteration matrix of many standard methods to be ill-conditioned \cite{brenanNumericalSolutionInitialValue1995}. To illustrate this, consider \eqref{eqn:bdf}, the $k$-step BDF method applied to \eqref{eqn:dae}. The vectors $\mathbf{x}_n$ and $\mathbf{z}_n$ are the predicted states at $t_n$, $\alpha_i$ are the BDF coefficients, and $h_n = t_n \!-\! t_{n\!-\!1}$. 

\vspace{-4mm}
\begin{subequations}
    \begin{align}
        \sum_{i=0}^{k} \alpha_{n,i} \mathbf{x}_{n-i} &= h_n\, \mathbf{f}(t_n, \mathbf{x}_n, \mathbf{z}_n) \\
        \mathbf{0} &= \mathbf{g}(t_n, \mathbf{x}_n, \mathbf{z}_n)
    \end{align}
    \label{eqn:bdf}
\end{subequations}
\vspace{-4mm}

To solve for $\mathbf{x}_n$ and $\mathbf{z}_n$, reformulate \eqref{eqn:bdf} as the nonlinear root-finding problem, \eqref{eqn:rootfinding}, and solve with Newton's method. Each iteration, $\nu$, of Newton requires solving the system of linear equations, \eqref{eqn:newton_linearsolve}, where $\Tilde{\mathbf{J}}$ is the \textit{Newton iteration matrix}.

\vspace{-4mm}
\begin{subequations}
    \begin{align}
        \Tilde{\mathbf{F}}(t_n, \mathbf{x}_n, \mathbf{z}_n) &=  \sum_{i=0}^{k}\alpha_{n,i} \mathbf{x}_{n-i} - h_n\, \mathbf{f}(t_n, \mathbf{x}_n, \mathbf{z}_n) = \mathbf{0} \\
        \Tilde{\mathbf{G}}(t_n, \mathbf{x}_n, \mathbf{z}_n) &= \mathbf{g}(t_n, \mathbf{x}_n, \mathbf{z}_n) = \mathbf{0}
    \end{align}
    \label{eqn:rootfinding}
\end{subequations}
\vspace{-4mm}
\begin{align}
    \Tilde{\mathbf{J}}(t_n, \mathbf{x}_n^{(\nu)}, \mathbf{z}_n^{(\nu)}) \!
    \begin{bmatrix}
        \mathbf{x}_n^{(\nu \!+\! 1)} \!-\! \mathbf{x}_n^{(\nu)} \\[1ex]
        \mathbf{z}_n^{(\nu \!+\! 1)} \!-\! \mathbf{z}_n^{(\nu)} \\
    \end{bmatrix} \!
    =
    \!-\!
    \begin{bmatrix}
        \Tilde{\mathbf{F}}(t_n, \mathbf{x}_n^{(\nu)}, \mathbf{z}_n^{(\nu)}) \\[1ex]
        \Tilde{\mathbf{G}}(t_n, \mathbf{x}_n^{(\nu)}, \mathbf{z}_n^{(\nu)}) 
    \end{bmatrix}\!
    \label{eqn:newton_linearsolve}
\end{align}
\vspace{1mm}

For a \textit{k}-step BDF method, the Newton iteration matrix is \eqref{eqn:newton_iteration_matrix}, evaluated at step $n$ and iteration $\nu$. When $\mathbf{g}_{\mathbf{z}}$ is singular, $\Tilde{\mathbf{J}}$ is ill-conditioned. Thus, small errors from evaluating $\Tilde{\mathbf{J}}$, $\Tilde{\mathbf{F}}$, or $\Tilde{\mathbf{G}}$, like those caused by finite precision rounding, will cause large errors in the solution, $\mathbf{x}_n^{(\nu\!+\!1)}$ and $\mathbf{z}_n^{(\nu\!+\!1)}$. Furthermore, $\Tilde{\mathbf{J}}$ becomes more ill-conditioned as $h_n$ decreases \cite{brenanNumericalSolutionInitialValue1995}. 

\vspace{-3mm}
\begin{align}
    \Tilde{\mathbf{J}}(t, \mathbf{x}, \mathbf{z}) =
    \begin{bmatrix}
        \frac{\partial \Tilde{\mathbf{F}}}{\partial \mathbf{x}} & \frac{\partial \Tilde{\mathbf{F}}}{\partial \mathbf{z}} \\[1ex]
        \frac{\partial \Tilde{\mathbf{G}}}{\partial \mathbf{x}} & \frac{\partial \Tilde{\mathbf{G}}}{\partial \mathbf{z}}
    \end{bmatrix}
    =
    \begin{bmatrix}
        {\small \alpha_0 \mathbf{I}} \!-\!h_n\mathbf{f}_{\mathbf{x}} & \ \  -h_n\mathbf{f}_{\mathbf{z}} \\[1ex]
        \mathbf{g}_{\mathbf{x}} & \ \  \mathbf{g}_{\mathbf{z}}
    \end{bmatrix}
    \label{eqn:newton_iteration_matrix}
\end{align}

Another challenge with integrating higher-index DAEs is that the analytical error estimates of $\mathbf{z}_n$ used for step size control do not tend to zero as $h_n$ decreases \cite{brenanNumericalSolutionInitialValue1995}. Therefore, the error estimates of $\mathbf{z}_n$ can force $h_n$ to decrease, causing $\Tilde{\mathbf{J}}$ to become more ill-conditioned, and eventually causing Newton failure \cite{ascherComputerMethodsOrdinary1998,brenanNumericalSolutionInitialValue1995}. Therefore, attempting to integrate an index-2 DAE with index-1 DAE solvers will likely lead to a Newton convergence error. Attributing this error to the DAE formulation is difficult because there are many possible reasons for convergence failure. Additionally, strategies to mitigate these issues do not work for every higher-index DAE system and thus must be tested on a case-by-case basis \cite{brenanNumericalSolutionInitialValue1995}.

\subsection{General index reduction}
\label{sec:practical_index_reduction}
\begin{figure}[!htp]
    \centering
    \includegraphics[width=8.3cm]{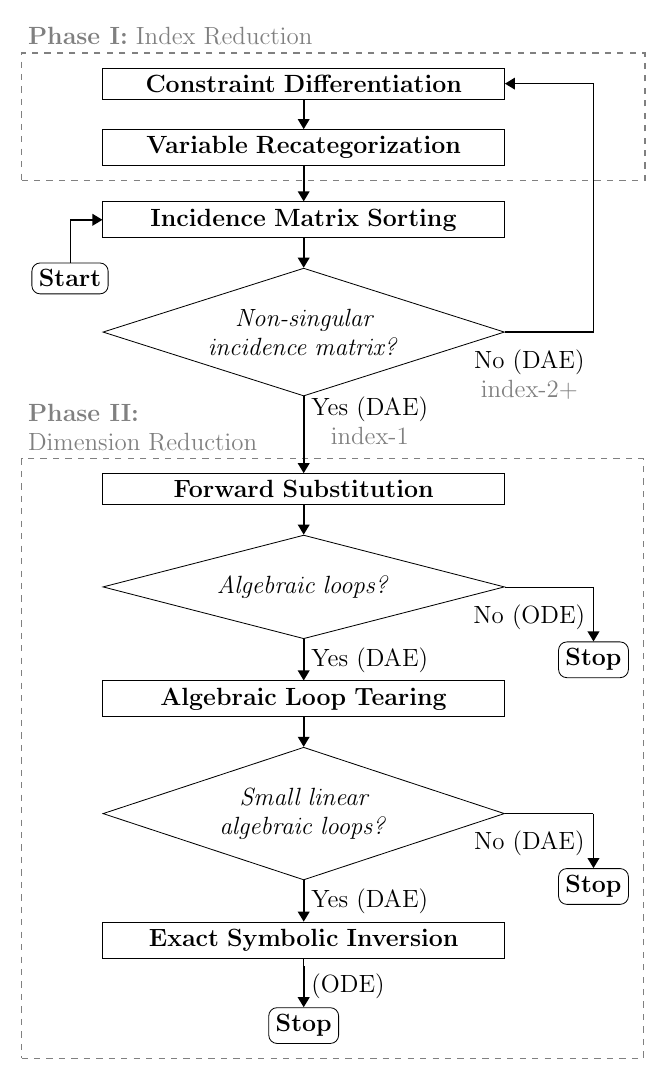}
    \caption{The typical algorithmic workflow in software tools for performing index reduction and dimension reduction on a DAE of unknown index. The DAE index is noted at each state of index reduction. The equation format (DAE or ODE) is noted at each stage of dimension reduction. }
    \label{fig:index_reduction_chart}
\end{figure}

Due to the difficulties of integrating higher-index DAEs directly, it is common to reformulate the original system into an index-1 DAE using index reduction. The process converts a DAE system of unknown index into a dynamically equivalent system that can be integrated with standard solvers. Fig.~\ref{fig:index_reduction_chart} shows the typical sequence of algorithms used in software tools. The specific implementation of each can vary.  
\subsubsection{Index Identification} To start, determine if index reduction is necessary with \textbf{incidence matrix sorting}. This algorithm attempts to convert the system incidence matrix, defined by \eqref{eqn:dae_incidence}, into lower-triangular (LT) or block-lower-triangular (BLT) form using row and column permutations. In LT form, each diagonal entry represents a causal relationship between an element of $\dot{\mathbf{x}}$ or $\mathbf{z}$ (column) and its defining equation (row). For higher-index DAEs, the singular incidence matrix cannot be permuted into LT or BLT form. In this case, sorting fails but reveals the $q$ constraint equations. Tarjan's algorithm is a common implementation of this step \cite{cellierContinuousSystemSimulation2006}.

\subsubsection{Index Reduction} \textbf{Phase I} begins if index reduction is necessary,  starting with \textbf{constraint differentiation}. All $q$ constraint equations are differentiated and added to the system, revealing the hidden constraint equations. Then, \textbf{variable recategorization} is performed to address the differential variable dependencies. For each constraint, one dependent differential variable is reclassified as algebraic. The time derivative of that variable, called a \textit{pseudo-derivative}, is also reclassified as algebraic. This converts both the constraint and hidden constraint into standard algebraic equations. Phase I reduces the index by one, adds $q$ equations and $2q$ algebraic variables, and removes $q$ differential variables. Phase I is repeated until the DAE is index-1, which is a valid stopping point. Pantelides algorithm is a common implementation of Phase I \cite{cellierContinuousSystemSimulation2006}.

\subsubsection{Dimension Reduction} \textbf{Phase II} (optional) attempts to reduce the dimension of the index-1 DAE\footnote{This is not technically the same as reducing the index of the Phase I output, which would involve differentiating all remaining algebraic equations.} to improve computational efficiency without approximations. If the sorted incidence matrix is LT, then all algebraic equations can be substituted into the differential equations using \textbf{forward substitution} resulting in an ODE. If the incidence matrix is in BLT form, only partial forward substitution is possible, leaving unsorted blocks that correspond to coupled algebraic loops. Algebraic loop dimensions can be reduced with \textbf{algebraic loop tearing}, which assumes one or more column-row causal relationships between \textit{tearing variables} and \textit{residual equations} to allow further incidence matrix sorting. If post-tearing algebraic loops are small and linear, they can be solved by \textbf{exact symbolic inversion}, reducing the system to an ODE. See Chapter 7 of \cite{cellierContinuousSystemSimulation2006} for details on most algorithms in Fig.~\ref{fig:index_reduction_chart}.

\subsection{Custom index reduction for power systems}
\label{sec:custom_index_reduction}
The general index reduction (GIR) method described above has been implemented in several general-purpose software tools. These algorithms are designed to handle a variety of systems, which requires them to symbolically represent and manipulate models at runtime using computer algebra. To perform time integration, the symbolic models are converted into numerical code using code generation. Unlike general-purpose tools, we are specifically interested in reducing the index of EMT-\textit{dq} models and do not need the flexibility of GIR. By combining domain knowledge with GIR, we can avoid the computational overhead of symbolic representation. EMT‑\textit{dq} models typically consist of interconnected subsystems, like transmission lines, transformers, and SGs, that are represented by dynamic equations. The follow sections describe why we can reduce the system index while remaining compatible with this modular approach. The result is two index-reduced topology-agnostic subsystem models, S1 and S1+S2, that can be implemented directly into numerical code. The derivations of both models are presented in Section~\ref{sec:subsystem_model_derivations}. 

\subsubsection{Index Identification} The goal of \textbf{incidence matrix sorting} is to determine the DAE index and identify the constraint equations. Since S1 and S2 are cutsets of inductors and/or current sources, we know the index and the constraints. According to Section~\ref{sec:model_formulation}, these structures cause the system to be an index-2 DAE with constraint equations from KCL. 

\subsubsection{Index Reduction}
The hidden constraints introduced during \textbf{constraint differentiation} are the derivatives of the known KCL constraints and thus can be differentiated analytically and included in the numerical subsystem model. \textbf{Variable recategorization} of the dependent differential variables and pseudo-derivatives can occur at the same time. Neither step requires knowledge of other subsystems. 

\subsubsection{Dimension Reduction}
After index reduction, there are algebraic loops corresponding to S1 and S1+S2. However, these are decoupled from the rest of the incidence matrix and thus can be solved without knowledge of other subsystems. For example, consider the block matrix, \eqref{eqn:irreducible}. 
\begin{align}
    B_{block} = \left[
    \begin{array}{ccc|c|c}
        D_1      & 0       & 0       & 0       & 0 \\
        L_{21}   & D_2     & 0       & 0       & 0 \\
        L_{31}   & L_{32}  & D_3     & 0       & 0 \\ \hline
        0        & 0       & 0       & D_4     & 0 \\ \hline
        L_{51}   & L_{52}  & L_{53}  & L_{54}  & D_5
    \end{array}
    \label{eqn:irreducible}
    \right]
\end{align}
It is a BLT sparsity pattern matrix partitioned into diagonal blocks, $D_i$, strictly lower blocks $L_{ij}$, and zero-valued upper blocks. When a diagonal block, like $D_4$, has zero-valued blocks to its left, its equations are structurally independent from the block rows above it. The variables in $D_4$ can be solved without any other matrices in $B_{block}$. 
This characteristic allows \textbf{forward substitution}, \textbf{algebraic loop tearing}, and \textbf{exact symbolic inversion} to be performed prior to runtime and implemented directly into the modular numerical models.

\section{Subsystem Model Derivations}
\label{sec:subsystem_model_derivations}
\noindent This section derives two custom index-reduced subsystem models by manually performing GIR on two minimal grid models that contain S1 and S1+S2, respectively. We demonstrate that the subsystem equations are not sensitive to the specific network topology and can be implemented in code as modular numerical models. Only the equations relevant to the final subsystem are explicitly shown. In both cases, the generator and network have the same per-unit base power. 

\subsection{Inverter to transformer}
\label{sec:inverter_to_transformer}
\begin{figure}[!ht]
    \centering
    \includegraphics[width=9cm]{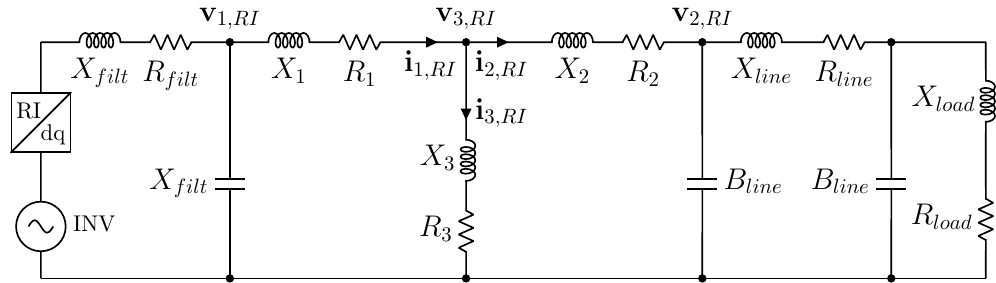}
    \caption{Schematic of a minimal grid featuring an inverter connected to a transformer, a $\pi$-section line, and a constant impedance load. All currents and voltages referenced in equations are labeled.}
    \label{fig:case1_lcfilter}
\end{figure}

Performing GIR on Fig.~\ref{fig:case1_lcfilter} derives the index-reduced subsystem model of S1, the transformer centered around $\mathbf{v}_{3,RI}$. The resulting model applies to any transformer connected on both sides to a bus whose complex voltage is a differential variable. In this case, one side is connected to an LC filter of an inverter and the other to a $\pi$-model transmission line. This model also applies to a transformer located between two $\pi$-model transmission lines.

We begin by defining the state variables, $\mathbf{x}$, and algebraic variables, $\mathbf{z}$, needed to define Fig.~\ref{fig:case1_lcfilter} as an EMT-\textit{dq} model. All variable categories are named, but only the variables needed for the derivation are labeled in Fig.~\ref{fig:case1_lcfilter} and listed below.

\vspace{-5mm}
\begin{align*}
    \mathbf{x} &= 
    \mathbf{x}_{\text{Inverter}} \,\cup\, 
    \mathbf{x}_{\text{Filter}} \,\cup\,
    \mathbf{x}_{\text{Transformer}} \,\cup\, 
    \mathbf{x}_{\text{Line}} \,\cup\, 
    \mathbf{x}_{\text{Load}} \\
    \mathbf{z} &= 
    \mathbf{z}_{\text{Inverter}} \,\cup\,
    \mathbf{z}_{\text{Filter}} \,\cup\,
    \mathbf{z}_{\text{Transformer}} \,\cup\, 
    \mathbf{z}_{\text{Line}} \,\cup\, 
    \mathbf{z}_{\text{Load}}
\end{align*}

where
\vspace{-3mm}
\begin{align*}
    \{
        \mathbf{v}_{1,RI} 
    \} 
    &\subseteq \mathbf{x}_{\text{Filter}}
    \\
    \{ 
        \mathbf{i}_{1,RI}, \ \mathbf{i}_{2,RI}, \ \mathbf{i}_{3,RI}
    \} 
    &\subseteq \mathbf{x}_{\text{Transformer}}\\
    \{
        \mathbf{v}_{3,RI}
    \}
    &\subseteq \mathbf{z}_{\text{Transformer}}\\
    \{
        \mathbf{v}_{2,RI}
    \} 
    &\subseteq \mathbf{x}_{\text{Line}} 
\end{align*}

Applying MNA to Fig.~\ref{fig:case1_lcfilter} causes the system to be an index-2 DAE, due to the cutset of inductors at $\mathbf{v}_{3,RI}$. The relevant subset of equations are \eqref{eqn:case1_index2}. 
\begin{subequations} 
    \begin{align}
        \frac{X_{1}}{\omega_0} \, \ddt{\mathbf{i}_{1,RI}} &= \mathbf{v}_{1,RI} - \mathbf{v}_{3,RI} - R_{1} \mathbf{i}_{1,RI} \pm X_{1} \mathbf{i}_{1,IR} \label{eqn:case1_index2_i1dq}\\
        \frac{X_{2}}{\omega_0} \, \ddt{\mathbf{i}_{2,RI}} &= \mathbf{v}_{3,RI} - \mathbf{v}_{2,RI} - R_{2} \mathbf{i}_{2,RI} \pm X_{2} \mathbf{i}_{2,IR} \label{eqn:case1_index2_i2dq}\\
        \frac{X_{3}}{\omega_0} \, \ddt{\mathbf{i}_{3,RI}} &= \mathbf{v}_{3,RI} - R_{3}\mathbf{i}_{3,RI} \pm X_{3} \mathbf{i}_{3,IR} \label{eqn:case1_index2_i3dq}\\
        0 &= \mathbf{i}_{1,RI} - \mathbf{i}_{2,RI} - \mathbf{i}_{3,RI} \label{eqn:case1_index2_kcldq}
    \end{align} \label{eqn:case1_index2}
\end{subequations}
where
\begin{align*}
    \{ \mathbf{i}_{1,RI}, \mathbf{i}_{2,RI}, \mathbf{i}_{3,RI}, \mathbf{v}_{1,RI}, \mathbf{v}_{2,RI} \} \subset \mathbf{x}, \quad
    \{ \mathbf{v}_{3,RI}\} \subset \mathbf{z} 
\end{align*}

The constraint equations are \eqref{eqn:case1_index2_kcldq}, since they are only a function of elements of $\mathbf{x}$. For constraint differentiation, add the derivatives of \eqref{eqn:case1_index2_kcldq} to the system and retain \eqref{eqn:case1_index2_kcldq}. For variable recategorization, select the shunt currents, $\mathbf{i}_{3,RI}$, as the dependent differential variables. Convert their derivatives, $\ddt{}\mathbf{i}_{3,RI}$, into pseudo-derivatives by renaming them to $\mathbf{di}_{3,RI}$. With the index-reduced transformer model, \eqref{eqn:case1_index1}, the full system becomes index-1 with a BLT incidence matrix.
\begin{subequations} 
    \begin{align}
        \frac{X_{1}}{\omega_0} \, \ddt{\mathbf{i}_{1,RI}} &= \mathbf{v}_{1,RI} - \mathbf{v}_{3,RI} - R_{1} \mathbf{i}_{1,RI} \pm X_{1} \mathbf{i}_{1,IR} \label{eqn:case1_index1_i1dq}\\
        \frac{X_{2}}{\omega_0} \, \ddt{\mathbf{i}_{2,RI}} &= \mathbf{v}_{3,RI} - \mathbf{v}_{2,RI} - R_{2} \mathbf{i}_{2,RI} \pm X_{2} \mathbf{i}_{2,IR} \label{eqn:case1_index1_i2dq}\\
        \frac{X_{3}}{\omega_0} \mathbf{di}_{3,RI} &= \mathbf{v}_{3,RI} - R_{3} \mathbf{i}_{3,RI} \pm X_{3} \mathbf{i}_{3,IR} \label{eqn:case1_index1_i3dq}\\
        0 &= \mathbf{i}_{1,RI} - \mathbf{i}_{2,RI} - \mathbf{i}_{3,RI} \label{eqn:case1_index1_kcldq} \\
        0 &= \ddt{\mathbf{i}_{1,RI}} - \ddt{\mathbf{i}_{2,RI}} - \mathbf{di}_{3,RI} \label{eqn:case1_index1_ddtkcldq}
    \end{align} \label{eqn:case1_index1}
\end{subequations}
\vspace{-5mm}

where 
\begin{align*}
    \{ \mathbf{i}_{1,RI}, \mathbf{i}_{2,RI}, \mathbf{v}_{1,RI}, \mathbf{v}_{2,RI} \} \subset \mathbf{x}, \quad
    \{ \mathbf{v}_{3,RI}, \mathbf{i}_{3,RI}, \mathbf{di}_{3,RI} \} \subset \mathbf{z}
\end{align*}

Finally, perform partial forward substitution of the index-1 BLT incidence matrix. Select $\mathbf{v}_{3,RI}$ as the tearing variables and \eqref{eqn:case1_index1_i3dq} as the residual equations. This reduces the algebraic loop to a 2x2 linear system which we invert directly to solve for $\mathbf{v}_{3,RI}$. The final subsystem model is \eqref{eqn:case1_index0}. Note that \eqref{eqn:case1_index0_v3dq} is separated for readability, but can be plugged into \eqref{eqn:case1_index0_i1dq}-\eqref{eqn:case1_index0_i2dq}.
\begin{subequations}
    \begin{align}
        &\mathbf{v}_{3,RI} \!=\! 
        \frac{
            \frac{\mathbf{v}_{1,RI}}{X_1}  \!+\! \frac{\mathbf{v}_{2,RI}}{X_2}  
            \!+\! \left(\frac{R_3}{X_3} \!-\! \frac{R_1}{X_1}\right)\mathbf{i}_{1,RI} 
            \!-\! \left(\frac{R_3}{X_3} \!-\! \frac{R_2}{X_2}\right) \mathbf{i}_{2,RI}
        }{
            \frac{1}{X_1} + \frac{1}{X_2} + \frac{1}{X_3}
        } 
        \label{eqn:case1_index0_v3dq} \\
        &\frac{X_{1}}{\omega_0} \, \ddt{\mathbf{i}_{1,RI}} = \mathbf{v}_{1,RI} - \mathbf{v}_{3,RI} - R_{1} \mathbf{i}_{1,RI} \pm X_{1} \mathbf{i}_{1,IR} \label{eqn:case1_index0_i1dq}\\
        &\frac{X_{2}}{\omega_0} \, \ddt{\mathbf{i}_{2,RI}} = \mathbf{v}_{3,RI} - \mathbf{v}_{2,RI} - R_{2} \mathbf{i}_{2,RI} \pm X_{2} \mathbf{i}_{2,IR} \label{eqn:case1_index0_i2dq}
    \end{align}
    \label{eqn:case1_index0}
\end{subequations}

The transformer subsystem model describes the dynamics of the independent interface currents, $\{\mathbf{i}_{1,RI}, \mathbf{i}_{2,RI}\}$ and correctly reflects the dependency of the magnetizing branch current and voltage, $\{\mathbf{i}_{3,RI}, \mathbf{v}_{3,RI}\}$. The equations are decoupled from everything except for the adjacent independent voltages, and thus can be implemented as a modular subsystem model. During this derivation, we made no additional approximations.

\subsection{Machine to transformer}
\begin{figure}[!ht]
    \centering
    \includegraphics[width=9cm]{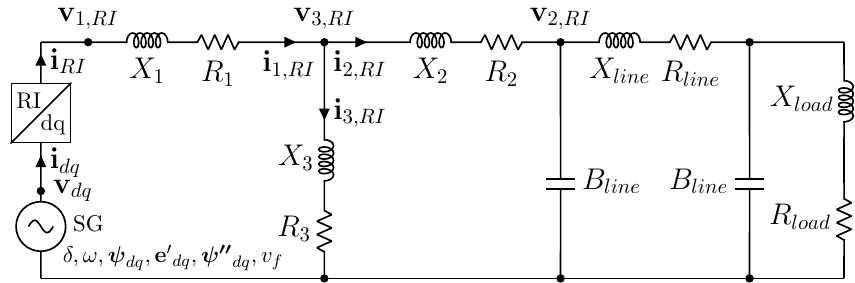}
    \caption{Schematic of a minimal grid featuring a SG connected to a transformer, a $\pi$-section line, and a constant impedance load.}
    \label{fig:case2_sg}
\end{figure}

Performing GIR on Fig.~\ref{fig:case2_sg} derives the index-reduced subsystem model of S1+S2. The transformer, S1, is centered around $\mathbf{v}_{3,RI}$, and the stator-transformer interface, S2, is at $\mathbf{v}_{1,RI}$. They are coupled through $\mathbf{i}_{RI}$, so the subsystem includes the machine, dynamic stator, and transformer. The transformer is also connected to a $\pi$-model transmission line. 

We begin by defining the state variables, $\mathbf{x}$, and algebraic variables, $\mathbf{z}$, needed to define Fig.~\ref{fig:case2_sg} as an EMT-\textit{dq} model. All variable categories are named, but only the variables needed for the derivation are labeled in Fig.~\ref{fig:case2_sg} and listed below.

\vspace{-5mm}
\begin{align*}
    \mathbf{x} &= 
    \mathbf{x}_{SG} \,\cup\, 
    \mathbf{x}_{Transformer} \,\cup\, 
    \mathbf{x}_{Line} \,\cup\, 
    \mathbf{x}_{Load} \\
    \mathbf{z} &= 
    \mathbf{z}_{SG} \,\cup\, 
    \mathbf{z}_{Transformer} \,\cup\, 
    \mathbf{z}_{Line} \,\cup\, 
    \mathbf{z}_{Load}
\end{align*}

where
\vspace{-3mm}
\begin{align*}
    \{
        \delta, \, \omega,  \, \boldsymbol{\psi}_{dq}, \, \mathbf{e'}_{dq}, \, \boldsymbol{\psi''}_{dq}, \, v_f
    \} 
    &\subseteq \mathbf{x}_{\text{SG}}
    \\
    \{ 
        \mathbf{i}_{dq}, \, \mathbf{v}_{dq}, \, \mathbf{i}_{RI}, \, \mathbf{v}_{1,RI} 
    \}
    &\subseteq \mathbf{z}_{\text{SG}}
    \\
    \{ 
        \mathbf{i}_{1,RI}, \, \mathbf{i}_{2,RI}, \, \mathbf{i}_{3,RI}
    \} 
    &\subseteq \mathbf{x}_{\text{Transformer}}\\
    \{
        \mathbf{v}_{3,RI}
    \}
    &\subseteq \mathbf{z}_{\text{Transformer}}\\
    \{
        \mathbf{v}_{2,RI}
    \} 
    &\subseteq \mathbf{x}_{\text{Line}} 
\end{align*}

Applying MNA to Fig.~\ref{fig:case2_sg} causes the system to be an index-2 DAE, due to the cutset of inductors at $\mathbf{v}_{3,RI}$ and the cutset of a current source (stator) and an inductor at $\mathbf{v}_{1,RI}$. The relevant subset of equations are \eqref{eqn:case1_index2} from Section~\ref{sec:inverter_to_transformer} and \eqref{eqn:case2_index2_kcl}-\eqref{eqn:case2_index2_i}.
\vspace{-4mm}
\begin{subequations} 
\begin{align}
    0 &= i_R - i_{1,R} 
    \label{eqn:case2_index2_kclR} \\
    0 &= i_I - i_{1,I} 
    \label{eqn:case2_index2_kclI}
\end{align} \label{eqn:case2_index2_kcl}
\end{subequations}
\vspace{-8mm}
\begin{subequations}
\begin{align}
    i_R &= i_d\sin{\delta} + i_q\cos{\delta} \label{eqn:case2_index2_iR} \\
    i_I &= -i_d\cos{\delta} + i_q\sin{\delta}  \label{eqn:case2_index2_iI} \\
    i_d &= \frac{1}{x''_d} \left( -\psi_d + \gamma_{d1}e'_q  + (1-\gamma_{d1})\psi''_d \right) \label{eqn:case2_index2_id}\\
    i_q &= \frac{1}{x''_q} \left( -\psi_q - \gamma_{q1}e'_d  + (1-\gamma_{q1})\psi''_q \right) \label{eqn:case2_index2_iq}
\end{align} \label{eqn:case2_index2_i}
\end{subequations}
The constraint equations associated with KCL at $\mathbf{v}_{3,RI}$ are \eqref{eqn:case2_index2_allkcls_v3d}-\eqref{eqn:case2_index2_allkcls_v3q}, which are identical to \eqref{eqn:case1_index2_kcldq}. The generator RF (dq) to network RF (RI) transformation is defined in \eqref{eqn:case2_index2_iR}-\eqref{eqn:case2_index2_iI}. The stator current is defined in \eqref{eqn:case2_index2_id}-\eqref{eqn:case2_index2_iq} in terms of machine states $\delta, \boldsymbol{\psi}_{dq},\mathbf{e'}_{dq},$ and $\boldsymbol{\psi''}_{dq}$. This example corresponds to the Sauer-Pai machine model defined in \cite{milanoPowerSystemModelling2010}, but the approach extends to any machine model with differentiable dynamic stator equations. Plugging \eqref{eqn:case2_index2_i} into \eqref{eqn:case2_index2_kcl} exposes the two constraint equations from KCL at $\mathbf{v}_{1,RI}$, shown in \eqref{eqn:case2_index2_allkcls_v1d}-\eqref{eqn:case2_index2_allkcls_v1q}. 
\begin{subequations}
    \begin{align}
        0 &= i_{1,R} - i_{2,R} - i_{3,R} \label{eqn:case2_index2_allkcls_v3d}\\
        0 &= i_{1,I} - i_{2,I} - i_{3,I} \label{eqn:case2_index2_allkcls_v3q}\\
        0 &= - i_{1,R} + \frac{1}{x''_d} \left( -\psi_d + \gamma_{d1}e'_q  + (1-\gamma_{d1})\psi''_d \right)\sin{\delta} \nonumber\\
        &\quad  + \frac{1}{x''_q} \left( -\psi_q - \gamma_{q1}e'_d  + (1-\gamma_{q1})\psi''_q \right)\cos{\delta}  \label{eqn:case2_index2_allkcls_v1d}\\
        0 &= - i_{1,I} -\frac{1}{x''_d} \left( -\psi_d + \gamma_{d1}e'_q  + (1-\gamma_{d1})\psi''_d \right)\cos{\delta} \nonumber\\
        &\quad + \frac{1}{x''_q} \left( -\psi_q - \gamma_{q1}e'_d  + (1-\gamma_{q1})\psi''_q \right)\sin{\delta} \label{eqn:case2_index2_allkcls_v1q}
    \end{align} \label{eqn:case2_index2_allkcls}
\end{subequations}

where 
\begin{align*}
    \{\mathbf{i}_{1,RI}, \mathbf{i}_{2,RI},\mathbf{i}_{3,RI}, \delta, \boldsymbol{\psi}_{dq},\mathbf{e'}_{dq},\boldsymbol{\psi''}_{dq} \} \subset \mathbf{x}
\end{align*}

Since these two pairs of constraint equations are coupled through $\mathbf{i}_{1,RI}$, they cannot be addressed independently. Differentiate \eqref{eqn:case2_index2_allkcls} to get the hidden constraints, \eqref{eqn:case2_index2_ddtKCL_v3d}-\eqref{eqn:case2_index2_ddtKCL_v1q}, and add them to the system. The derivatives of \eqref{eqn:case2_index2_i} are separated for clarity.
\begin{subequations}
    \begin{align}
        0 &= \ddt{i_{1,R}} - \ddt{i_{2,R}} - \ddt{i_{3,R}} \label{eqn:case2_index2_ddtKCL_v3d} \\
        0 &= \ddt{i_{1,I}} - \ddt{i_{2,I}} - \ddt{i_{3,I}} \label{eqn:case2_index2_ddtKCL_v3q} \\
        0 &= \ddt{i_R} - \ddt{i_{1,R}} \label{eqn:case2_index2_ddtKCL_v1d}\\
        0 &= \ddt{i_I} - \ddt{i_{1,I}} \label{eqn:case2_index2_ddtKCL_v1q} \\
        \ddt{i_R} &= i_d\cos{\delta} + \ddt{i_d}\sin{\delta} - i_q\sin{\delta} + \ddt{i_q}\cos{\delta} \\
        \ddt{i_I} &= i_d\sin{\delta}  - \ddt{i_d}\cos{\delta} + i_q\cos{\delta} + \ddt{i_q}\sin{\delta} \\
        x''_d \, \ddt{i_d} &= -\ddt{\psi_d} + \gamma_{d1}\ddt{e'_q}  + (1-\gamma_{d1})\ddt{\psi''_d} \\
        x''_q \, \ddt{i_q} &= -\ddt{\psi_q} - \gamma_{q1}\ddt{e'_d}  + (1-\gamma_{q1})\ddt{\psi''_q} 
    \end{align} \label{eqn:case2_index2_ddtKCL}
\end{subequations}

For variable recategorization, select $\mathbf{i}_{3,RI}$ and $\mathbf{i}_{1,RI}$ as the four dependent differential variables. Rename the pseudo-derivatives, $\ddt{}\mathbf{i}_{3,RI}$ and $\ddt{}\mathbf{i}_{1,RI}$, to $\mathbf{di}_{3,RI}$ and $\mathbf{di}_{1,RI}$. Rename the derivatives in the corresponding transformer equations as well. With this index-reduced model, the system is index-1 with a BLT incidence matrix (not shown).

For the optional dimension reduction, perform partial forward substitution on the incidence matrix. Then, select $\mathbf{v}_{3,RI}$ and $\mathbf{v}_{1,RI}$ as the four tearing variables and \eqref{eqn:case1_index1_i1dq} and \eqref{eqn:case1_index1_i3dq} as the residual equations. For this example, $\ddt{}\mathbf{i}_{1,RI}$ becomes $\mathbf{di}_{1,RI}$ in \eqref{eqn:case1_index1_i1dq}. Similar to Section~\ref{sec:inverter_to_transformer}, tearing reveals a linear system of equations where the unknowns are the tearing variables, shown in \eqref{eqn:case2_tearing}. This case results in a 4x4 matrix since the two sets of constraint equations are coupled. The upper left corner the linear system matrix is diagonal, so we can solve this subsystem directly using the Schur complement. Thus, we can find analytical expressions for $\mathbf{v}_{3,RI}$ and $\mathbf{v}_{1,RI}$ to use in the final subsystem model. 
\begin{subequations}
    \begin{align}
        &\begin{bmatrix}
            - \frac{1}{X_1} & 0 & \frac{1}{X_1} + \Phi_{dq} & \Theta  \\
            0 & - \frac{1}{X_1} & \Theta & \frac{1}{X_1} + \Phi_{qd}  \\
            \frac{1}{X_3} + \frac{1}{X_2} & 0 & \Phi_{dq} & \Theta  \\
             0 & \frac{1}{X_3} + \frac{1}{X_2} & \Theta &  \Phi_{qd}
        \end{bmatrix}
        \begin{bmatrix}
            v_{3,R} \\ v_{3,I} \\ v_{1,R} \\ v_{1,I}
        \end{bmatrix}
        \nonumber \\ &=
        \begin{bmatrix}
            \frac{\beta_R}{\omega_0} + \frac{R_1}{X_1} i_{1,R} - i_{1,I} \\
            \frac{\beta_I}{\omega_0} + \frac{R_1}{X_1} i_{1,I} + i_{1,R} \\
            \frac{\beta_R}{\omega_0} + \frac{1}{X_2}v_{2,R}  + \frac{R_2}{X_2} i_{2,R} + \frac{R_3}{X_3} i_{3,R} - i_{2,I} - i_{3,I} \\
            \frac{\beta_I}{\omega_0} + \frac{1}{X_2}v_{2,I} + \frac{R_2}{X_2} i_{2,I} + \frac{R_3}{X_3} i_{3,I} + i_{2,R} + i_{3,R}
        \end{bmatrix}
    \end{align}
    
    where
    \begin{align}
        \Phi_{dq} &= \frac{1}{x''_d}\sin^2{\delta} + \frac{1}{x''_q}\cos^2{\delta} \\
        \Phi_{qd} &= \frac{1}{x''_d}\cos^2{\delta} + \frac{1}{x''_q}\sin^2{\delta} \\
        \Theta &= \left(-\frac{1}{x''_d} + \frac{1}{x''_q}\right) \sin{\delta}\cos{\delta} \\
        \beta_R &= 
            \left[ i_d + \frac{\omega_0}{x''_q} \left(\Gamma_{q1} - r_a i_q + \omega \psi_d \right) \right] \cos{\delta} \nonumber \\
            & + \left[ - i_q + \frac{\omega_0}{x''_d} \left(\Gamma_{d1} - r_a i_d - \omega \psi_q \right) \right] \sin{\delta} \\
        \beta_I &= 
            \left[ i_d + \frac{\omega_0}{x''_q} \left(\Gamma_{q1} - r_a i_q + \omega \psi_d \right) \right] \sin{\delta} \nonumber \\
            & + \left[ i_q - \frac{\omega_0}{x''_d} \left(\Gamma_{d1} - r_a i_d - \omega \psi_q \right) \right] \cos{\delta}  \\
        \Gamma_{d1} &= \frac{1}{\omega_0} \left(\gamma_{d1}\ddt{e'_q}  + (1-\gamma_{d1})\ddt{\psi''_d}\right) \\
        \Gamma_{q1} &= \frac{1}{\omega_0} \left(- \gamma_{q1}\ddt{e'_d}  + (1-\gamma_{q1})\ddt{\psi''_q} \right)
    \end{align} \label{eqn:case2_tearing}
\end{subequations}

The index-reduced model of a Sauer-Pai machine, dynamic stator, and transformer is shown in \eqref{eqn:case2_final}. The pre-substitution form is shown for readability, but it can be expressed as an ODE using forward substitution. 

\begin{subequations}
\begin{align}
    i_d &= \frac{1}{x''_d} \left( -\psi_d + \gamma_{d1}e'_q  + (1-\gamma_{d1})\psi''_d \right) \\
    i_q &= \frac{1}{x''_q} \left( -\psi_q - \gamma_{q1}e'_d  + (1-\gamma_{q1})\psi''_q \right) \\
    i_{1,R} &= i_d\sin{\delta} + i_q\cos{\delta} \\
    i_{1,I} &= -i_d\cos{\delta} + i_q\sin{\delta} \\
    i_{3,R} &= i_{1,R} - i_{2,R} \\
    i_{3,I} &= i_{1,I} - i_{2,I} \\
    T'_{d0}\ddt{e'_q} &= - (x_d - x'_d)\left(\gamma_{d1}i_d -\gamma_{d2}\psi''_d + \gamma_{d2} e'_q\right) \nonumber \\ & \quad \quad -e'_q + v_f \\
    T'_{q0}\ddt{e'_d} &= (x_q \!-\! x'_q)\left(\gamma_{q1}i_q \!-\! \gamma_{q2}\psi''_q \!-\! \gamma_{d2} e'_d\right) \!-\! e'_d 
    \end{align}    
    \begin{align}  
    T''_{d0}\ddt{\psi''_d} &= -\psi''_d + e'_q - (x'_d - x_l)i_d\\
    T''_{q0}\ddt{\psi''_q} &= -\psi''_q - e'_d - (x'_q - x_l)i_q \\
    [
        v_{3,R} &\,\,\, v_{3,I} \,\,\, v_{1,R} \,\,\, v_{1,I}
    ]^T = \mathrm{soln.\ to\ \eqref{eqn:case2_tearing}} \\ 
    v_d &= v_{1,R}\sin{\delta} - v_{1,I}\cos{\delta} \\
    v_q &= v_{1,R}\cos{\delta} + v_{1,I}\sin{\delta} \\
    \frac{1}{\omega_0} \, \ddt{\psi_d}  &= r_a i_d + \omega \psi_q + v_d \\
    \frac{1}{\omega_0} \, \ddt{\psi_q}  &= r_a i_q - \omega \psi_d + v_q \\
    \frac{X_{2}}{\omega_0} \, \ddt{i_{2,R}} &= v_{3,R} - v_{2,R} - R_{2} i_{2,R} + X_{2} i_{2,I} \\
    \frac{X_{2}}{\omega_0} \, \ddt{i_{2,I}} &= v_{3,I} - v_{2,I} - R_{2} i_{2,I} - X_{2} i_{2,R}
\end{align} \label{eqn:case2_final}
\end{subequations}

The subsystem model, \eqref{eqn:case2_final}, describes the dynamics of the combined S1+S2 subsystem and correctly reflects the dependency of both the low-side and magnetizing branch of the transformer. The equations are decoupled from everything except for the high-side independent voltage, and thus can be implemented as a modular subsystem model. During this derivation, we made no additional approximations.

In the remaining sections, custom index reduction (CIR) refers to the process of constructing an numerical EMT-\textit{dq} model with one or both of these derived subsystem models.

\section{Experimental Design}
\label{sec:experimental_design}
\noindent The objective of these experiments is to evaluate the performance of constructing a numerical index-reduced EMT-\textit{dq} model using CIR versus GIR across varying system sizes. We also verify the equivalence of CIR and GIR. This section details the methodology used to obtain the results in Section~\ref{sec:experimental_results}.

\subsection{Power system models}
The experiments use eight EMT-\textit{dq} models of increasing size. Each model is built from of $n$ instances of a 9-bus base system. The instances are connected together with additional lines chosen as a random graph. The load and network topology of each 9-bus instance are based on the WSCC 9-bus system \cite{sauerPowerSystemDynamics2017}. The network and generators are modeled dynamically using a balanced EMT-\textit{dq} framework. Since network-wide EMT simulations are most relevant for grids with inverters, we chose a two to one ratio of SGs to inverters within each 9-bus instance. The dynamic SG models include a 6th-order Sauer-Pai machine, dynamic stator, and Type I automatic voltage regulator \cite{milanoPowerSystemModelling2010}. The dynamic grid-forming inverter (GFM) model includes voltage inner control, active and reactive droop outer control, and an LC filter. Table~\ref{tab:scaling_cases} provides details about the eight cases used for this analysis. We built large models from stable substructures to improve the likelihood of initialization and enable the evaluation of the proposed method. Connecting smaller systems introduces non-realistic repeated structures, but test systems with dynamic networks and over 1000 buses are not widely available.

\begin{table}[!ht]
    \caption{Topological details of scaled systems}
    \label{tab:scaling_cases}
    \renewcommand{\arraystretch}{1.2}
    \setlength{\tabcolsep}{4pt} 
    \centering
    \begin{tabular}{|c|c|c|c|c|c|c|c|c|}
        \hline
        \textbf{Case} & \textbf{C1} & \textbf{C2} & \textbf{C3} & \textbf{C4} & \textbf{C5} & \textbf{C6} & \textbf{C7} & \textbf{C8}\\
        \hline 
        Buses & 9 & 18 & 36 & 72 & 144 & 288 & 576 & 1152\\
        \hline
        9-bus instances & 1 & 2 & 4 & 8 & 16 & 32 & 64 & 128\\
        \hline
        State variables & 68 & 139 & 283 & 567 & 1135 & 2271 & 4543 & 9087 \\
        \hline
        SGs & 2 & 4 & 8 & 16 & 32 & 64 & 128 & 256\\
        \hline
        Inverters & 1 & 2 & 4 & 8 & 16 & 32 & 64 & 128\\
        \hline
        Lines & 6 & 13 & 28 & 56 & 112 & 224 & 448 & 896\\
        \hline
        Transformers & 3 & 6 & 12 & 24 & 48 & 96 & 192 & 384\\
        \hline
    \end{tabular}
\end{table}
\vspace{-5mm}

\subsection{Experiments}
The primary focus of these experiments is performance improvements associated with numerical model construction. However, the resulting models are also integrated to validate equivalency and to place the model-building process in the context of the full simulation.

For the \textit{model build}, we build each of the eight cases five times and measure the total allocated memory, maximum resident set size, and wall clock runtime. Both implementations result in a Julia object called \texttt{ODEProblem}, a numerical model that can be solved with time-integration methods available in DifferentialEquations.jl \cite{rackauckasDifferentialEquationsjlPerformantFeatureRich2017}. Since Julia uses just-in-time compilation, we throw out the first run of each experiment since it can be skewed by compilation overhead. 

For the \textit{integration}, we integrate each of the eight cases five times and measure runtime. For all integration runs, a 20\% load step is applied at bus 8 of the original 9-bus instance. We use \texttt{Rodas5P}, a 5th-order Rosenbrock method, with a 4th-order interpolant to allow direct comparison of specific time points. To validate, we compare the trajectories of CIR against that of GIR, using infinity norms of the trajectories and eigenvalues. The infinity norm of a vector, $\infnorm{*}$, is the maximum absolute value. The first run of each experiment is again ignored due to just-in-time compilation. 

\subsection{Software and hardware}
All experiments were implemented in Julia and performed on the Alpine cluster at University of Colorado Boulder \cite{universityofcoloradoboulderresearchcomputingAlpine2023} using a single AMD Milan (x86\_64) CPU node. We used up to 64 cores for  model construction, each with 3.75 GB RAM (240 GB RAM total), and up to 24 cores for numerical integration, each with 12 GB RAM (384 GB total). GIR was implemented using ModelingToolkit.jl \cite{maModelingToolkitComposableGraph2021} and CIR was implemented within PowerSimulationDynamics.jl \cite{laraPowerSimulationsDynamicsjlOpenSource2024}. The code to recreate the experimental results can be found in \cite{majeauDAEIndexReductionforEMTModels2026}.

\section{Experimental Results}
\label{sec:experimental_results}

\subsection{Equivalence verification}
To validate the models generated by CIR, we compare simulation results for a specific perturbation and eigenvalues of the initial condition. Table~\ref{tab:diff_inf_norms} shows the difference metrics used for these comparisons. Subscripts G and C refer to GIR and CIR, respectively. The vector $\boldsymbol{x}$ is the value of state $x$ from $t_{start}$ to $t_{end}$. The vector $\boldsymbol{\lambda}$ is the eigenvalues of the pre-perturbation Jacobian evaluated at the initial condition.  

Column 1 of Table~\ref{tab:diff_inf_norms} shows the largest simulation error of any state variable at any time step. For example, when C8 is perturbed, the trajectories generated by the two methods never differ by more than 1.8e-04 pu for any of the 9087 state variables. Column 2 of Table~\ref{tab:diff_inf_norms} shows the mean infinity norm across all state variables. Column 3 of Table~\ref{tab:diff_inf_norms} compares the eigenvalues of the Jacobian evaluated at the initial condition. The difference norms are very small, indicating that the CIR and GIR index-reduced models are sufficiently equivalent. 

\vspace{-3mm}
\begin{table}[!htp]
    \caption{Difference between GIR-based and CIR-based simulations}
    \label{tab:diff_inf_norms}
    \renewcommand{\arraystretch}{1.2}
    \setlength{\tabcolsep}{4pt} 
    \centering
    \begin{tabular}{|c|c|c|c|}
        \hline
        \textbf{Buses} & \textbf{Max} $\infnorm{\boldsymbol{x}_G - \boldsymbol{x}_C}$ & \textbf{Mean} $\infnorm{\boldsymbol{x}_G - \boldsymbol{x}_C}$ & $ \infnorm{\lambda_G -\lambda_C} $ \\
        \hline 
        9 & 2.1e-05 & 3.2e-06 & 1.6e-08 \\
        \hline
        18 & 1.3e-05 & 1.5e-06 & 4.7e-09 \\
        \hline
        36 & 1.4e-05 & 1.6e-06 & 4.7e-09 \\
        \hline
        72 & 3.3e-05 & 1.4e-06 & 5.8e-09 \\
        \hline
        144 & 3.6e-05 & 1.2e-06 & 5.9e-09 \\
        \hline
        288 & 7.7e-05 & 9.8e-07 & 1.9e-08 \\
        \hline
        576 & 1.4e-04 & 8.6e-07 & 1.0e-08 \\
        \hline
        1152 & 1.8e-04 & 8.1e-07 & 1.1e-08 \\
        \hline
    \end{tabular}
\end{table}

\begin{figure}[!htp]
    \includegraphics[width=8.8cm,]{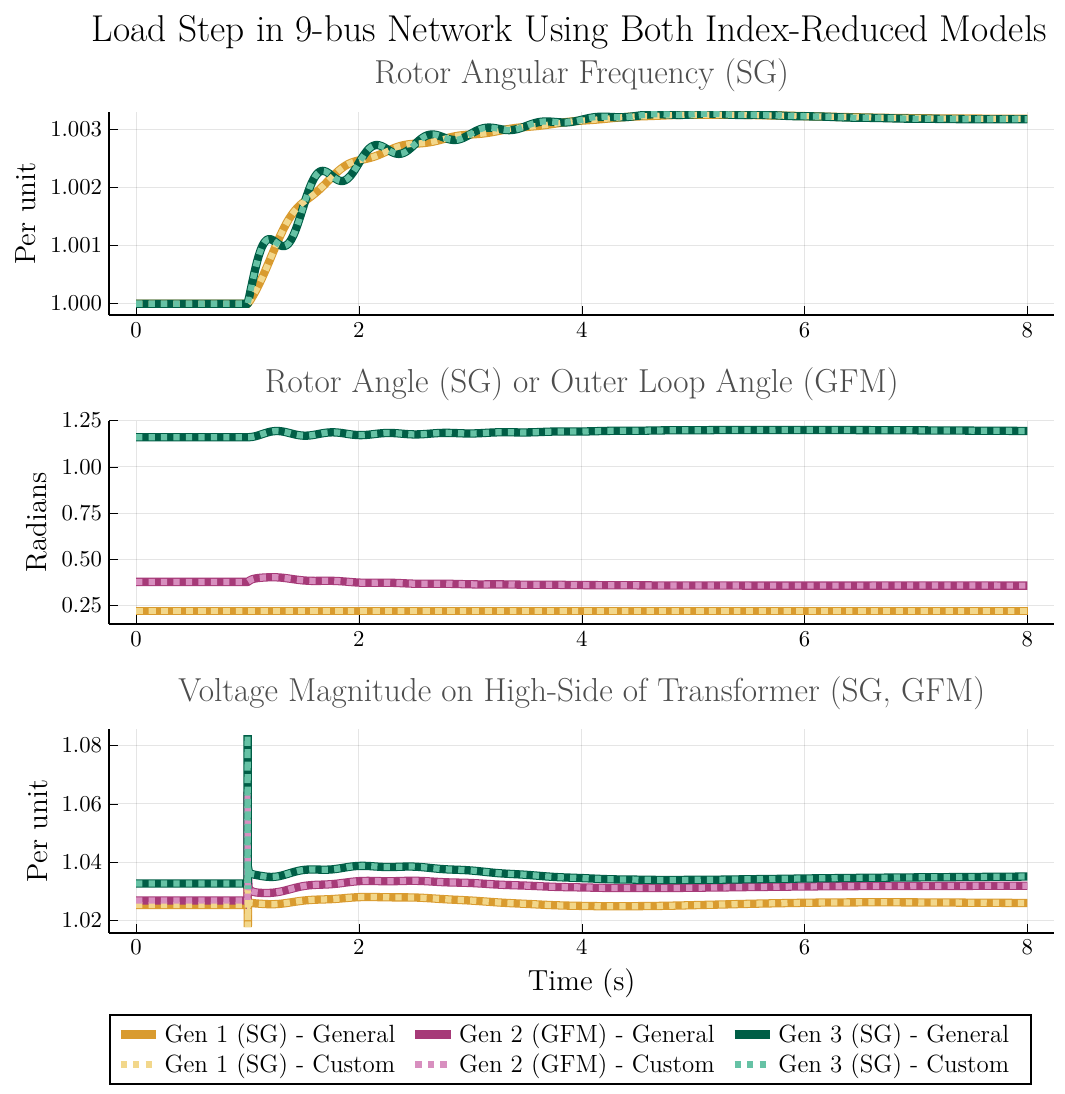}
    \caption{Comparison of the GIR and CIR index-reduced 9-bus models for a 20\% load step at Bus 8. Only a subset of the 68 state variables are shown for readability. Metrics in Table~\ref{tab:diff_inf_norms} are based on all variables.}
    \label{fig:validation_trajectory_samples_6Bus_2sg1inv}
\end{figure}

Fig.~\ref{fig:validation_trajectory_samples_6Bus_2sg1inv} shows a selection of the simulation results summarized in Table~\ref{tab:diff_inf_norms}. Only a subset of state variables from the 9-bus system are shown for readability. As suggested by the validation metrics in Table~\ref{tab:diff_inf_norms}, there are no perceptible differences between the two methods in these plots. 

\subsection{Performance comparison of model build}
This section considers the relative computation performance of building a numerical model using CIR and GIR. Fig.~\ref{fig:plot_performance_malloc}, Fig.~\ref{fig:plot_performance_maxrss}, and Fig.~\ref{fig:plot_performance_runtime} illustrate the scaling of several performance metrics for both approaches.

\subsubsection{Memory scaling analysis}
\begin{figure}[!ht]
    \centering
    \includegraphics[width=9cm,]{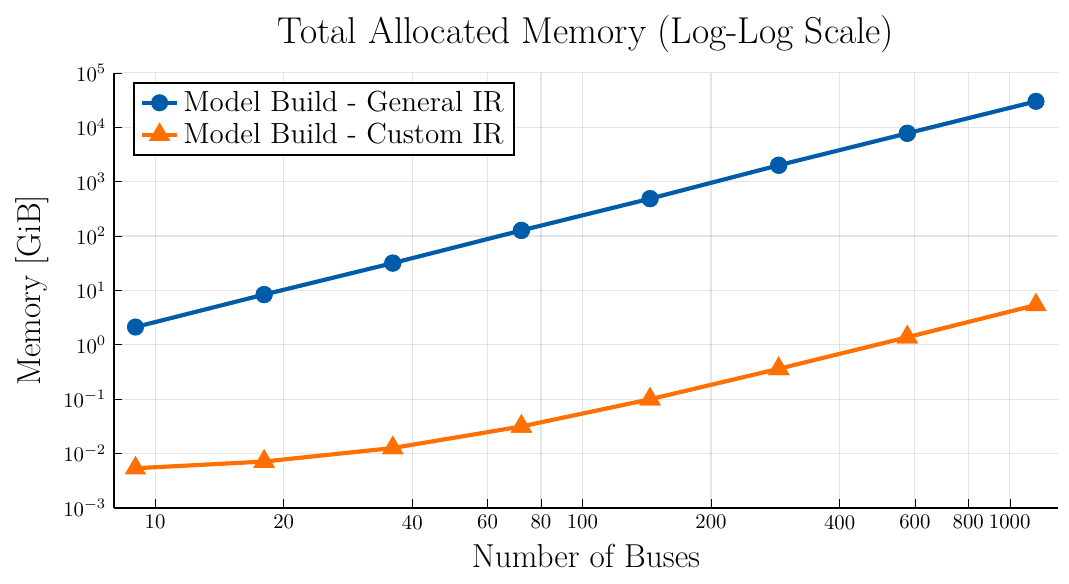}
    \caption{Memory allocated during build of \texttt{ODEProblem} assuming the initial condition is already known.}
    \label{fig:plot_performance_malloc}
\end{figure}

\begin{figure}[!ht]
    \centering
    \includegraphics[width=9cm,]{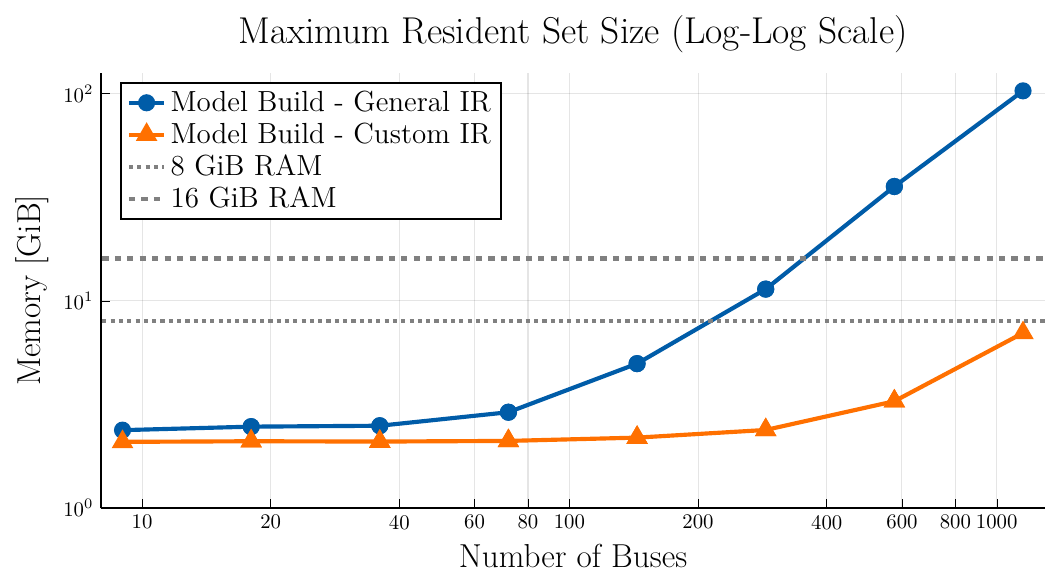}
    \caption{Highest concurrent memory usage during build of \texttt{ODEProblem} assuming the initial condition is already known.}
    \label{fig:plot_performance_maxrss}
\end{figure}

Fig.~\ref{fig:plot_performance_malloc} and Fig.~\ref{fig:plot_performance_maxrss}, show that CIR is less memory intensive than GIR by several orders of magnitude. Fig.~\ref{fig:plot_performance_malloc} states that the total memory allocated throughout the development of a numerical model is $10^{2.6}$ to $10^{3.7}$ times larger for GIR. Both approaches scale roughly as a power law. Fig.~\ref{fig:plot_performance_maxrss} shows that the maximum resident set size (RSS)\footnote{RSS is the RAM usage at a single point in time. Maximum RSS is a reasonable indicator of the RAM required to run a process.} during the development of the numerical model is up to 10 times larger for GIR. This metric is only sensitive to system size for the larger models, once the memory needed to build the system exceeds the program overhead. The horizontal lines in Fig.~\ref{fig:plot_performance_maxrss} show two RAM sizes for context. A standard laptop with 16 GiB\footnote{RAM is typically defined in GiB ($2^{30}$ bytes), not GB ($10^9$ bytes).} of RAM would likely experience out-of-memory errors when building the 288-bus model with GIR, assuming some unavoidable operating system overhead. For both approaches, Fig.~\ref{fig:plot_performance_malloc} indicates that total allocated memory scales as approximately $O(n^{1.5})$ where $n$ is the number of buses. Precise scaling is harder to quantify for the maximum RSS, because the program overhead is not surpassed until over 60 buses (GIR) and over 200 buses (CIR). The last few data points of GIR and CIR in Fig.~\ref{fig:plot_performance_maxrss} suggest approximately $O(n^{1.5})$ and $O(n)$ scaling, respectively. 

\subsubsection{Runtime scaling analysis}
\begin{figure}[!ht]
    \centering
    \includegraphics[width=9cm,]{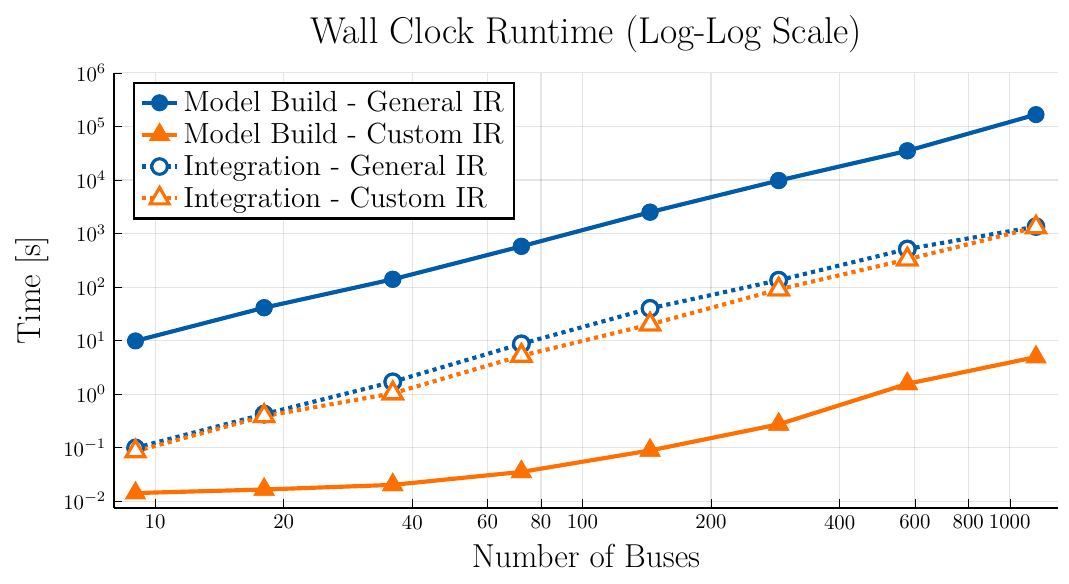}
    \caption{Wall clock time to build an \texttt{ODEProblem} assuming the initial condition is already known. The simulation runtime is included for reference.}
    \label{fig:plot_performance_runtime}
\end{figure}

Fig.~\ref{fig:plot_performance_runtime} compares wall clock time between the two approaches. Building a numerical model with CIR is $10^{2.8}$ to $10^{4.5}$ times faster than GIR, shown by the solid lines. We are primarily interested in comparing the model build performance, but integration runtime is included for reference, shown by the dashed lines. CIR improves the runtime so drastically that the overall bottleneck of the CIR-based simulation is the time integration, not the model build like with GIR. For both approaches, the runtime scales as approximately $O(n^{1.5})$ where $n$ is the number of buses.

\subsubsection{Discussion}
These experiments measure the computational complexity of building an index-reduced EMT-\textit{dq} model that can be integrated by standard solvers. With GIR, the starting point is a higher-index DAE in symbolic form. With CIR, the starting point is a set of modular index-reduced numerical models. Building and manipulating a symbolic model is more memory intensive than performing calculations with numerical values because symbolic models represent expressions as structured objects rather than numbers. This is the primary reason for the significant memory discrepancies. Generating numerical code from a symbolic model is time intensive because the system must traverse the symbolic model and construct corresponding code that expresses the same system of equations and can be compiled for use by numerical solvers. This the primary reason for the significant runtime discrepancies between the two methods.

\section{Conclusion}
\label{sec:conclusion}

\noindent In this paper, we present two modular index-reduced subsystem models that allow large network-wide EMT-\textit{dq} models to be built and integrated with standard solvers, without incurring the significant computational costs associated with symbolic representation. Building an index-reduced EMT-\textit{dq} model with these subsystem models is several orders of magnitude faster and less memory intensive than the general index reduction approach. These runtime improvements cause the computational bottleneck of generating EMT-\textit{dq} simulations to be time integration, not model construction. Future work will include evaluating the performance of these index-reduced models for a range of integration methods and developing more realistic EMT-\textit{dq} benchmark models.

\section*{Acknowledgments}
{
\noindent This material is based upon work supported by the U.S. Department of Energy, Office of Science, Office of Advanced Scientific Computing Research, Department of Energy Computational Science Graduate Fellowship under Award Number DE-SC0024386. This report was prepared as an account of work sponsored by an agency of the United States Government. Neither the United States Government nor any agency thereof, nor any of their employees, makes any warranty, express or implied, or assumes any legal liability or responsibility for the accuracy, completeness, or usefulness of any information, apparatus, product, or process disclosed, or represents that its use would not infringe privately owned rights. Reference herein to any specific commercial product, process, or service by trade name, trademark, manufacturer, or otherwise does not necessarily constitute or imply its endorsement, recommendation, or favoring by the United States Government or any agency thereof. The views and opinions of authors expressed herein do not necessarily state or reflect those of the United States Government or any agency thereof.

This work was authored in part by the National Laboratory of the Rockies for the U.S. Department of Energy (DOE), operated under Contract No. DE-AC36-08GO28308. Funding provided by applicable Department of Energy office and program office, e.g., U.S. Department of Energy Office of Electricity and Grid Deployment office. The views expressed in the article do not necessarily represent the views of the DOE or the U.S. Government. The U.S. Government retains and the publisher, by accepting the article for publication, acknowledges that the U.S. Government retains a nonexclusive, paid-up, irrevocable, worldwide license to publish or reproduce the published form of this work, or allow others to do so, for U.S. Government purposes.

This work utilized the Alpine high performance computing resource at the University of Colorado Boulder. Alpine is jointly funded by the University of Colorado Boulder, the University of Colorado Anschutz, Colorado State University, and the National Science Foundation (award 2201538).
}

\bibliographystyle{IEEEtran}
\bibliography{references}

\end{document}